\documentclass[aps,prd,twocolumn,longbibliography,reprint,groupedaddress,nofootinbib]{revtex4-1} 
\usepackage{amsmath,amsfonts,amsthm,amssymb,mathrsfs}
\usepackage[utf8]{inputenc}
\usepackage{natbib}
\usepackage[dvips]{graphics,graphicx}
\usepackage[usenames,dvipsnames]{color}
\definecolor{darkblue}{RGB}{0,0,196}
\definecolor{darkgreen}{RGB}{0,120,0}
\usepackage[colorlinks=true,linktocpage=true,linkcolor=darkblue,citecolor=darkblue,
urlcolor=darkblue]{hyperref}
\usepackage[normalem]{ulem}
\usepackage{cancel}
\usepackage{multirow}
\usepackage{longtable}
\usepackage{textcomp}
\usepackage{subfigure}

\begin{document}

\preprint{}

\title{Rotating Bose-Einstein Condensate Stars at finite temperature
}

\author{P. S. Aswathi}

\author{P. S. Keerthi}

\author{O. P. Jyothilakshmi}

\author{Lakshmi J. Naik}

\author{V. Sreekanth}
\email{v\_sreekanth@cb.amrita.edu}

\affiliation{Department of Sciences, Amrita School of Physical Sciences, Coimbatore, Amrita Vishwa Vidyapeetham, India}

\date{\today}
\begin{abstract}

We study the effect of temperature on the global properties of static and slowly rotating self-gravitating Bose-Einstein condensate (BEC) stars within general relativity.
We employ a recently developed temperature dependent BEC equation of state (EoS) to describe the stellar matter by assuming that the condensate can be described by a non-relativistic EoS.
Stellar profiles are obtained using general relativistic Hartle-Thorne slow rotation approximation equations.
We find that with increasing temperatures mass-radius values are found to be decreasing for the static and rotating cases; though presence of temperature supports high mass values at lower central densities.
Countering effects of rotation and temperature on the BEC stellar structure have been analysed and quantified. We report that inclusion of temperature has significant effect on the rotating stellar profiles but negligible effect on the maximum mass, as in the case of static system.
We have also studied the effect of EoS parameters- boson mass and strength of the self-interaction on global properties of static and rotating BEC stars, in presence of temperature.

\end{abstract}


\maketitle
\section{Introduction}
\label{intro}

The assumption that fundamental scalar fields exist in nature is supported by the detection of scalar particle Higg's boson at the Large Hadron Collider, CERN~\cite{ATLAS:2012yve,CMS:2012qbp}. Such scalar fields bound together by their self-generated gravitational interaction can form compact astrophysical objects known as boson stars (see Refs.~\cite{Jetzer:1991jr,Liddle:1992fmk,Schunck:2003kk,Liebling:2012fv} for comprehensive reviews on this topic).
It is well known that the scalar fields represent identical bosonic particles that can occupy the same quantum ground state at very low temperatures, forming a Bose-Einstein condensate (BEC). The BEC state, which was predicted by Bose and Einstein, was first experimentally realized by confining ultra-cold gas of rubidium atoms in a magnetic trap (see Refs.~\cite{BEC_nature_2002,BEC_nature_2020} for reviews on BECs).
Experiments on such trapped dilute Bose gases suggest a phase transition of atoms to the Bose condensed phase, with the particles occupying a coherent state~\cite{Dalfovo_1999,pitaevskii_2003,Pethick:2002}. Similarly, a coherent massive object such as boson star can also be realized within astrophysical scales formed of BECs, confined by the self-generated gravitational interaction of bosonic particles. Further, the analysis of Ref.~\cite{ODell_2000} with intense off-resonant laser beams also suggest the possibility of mimicking such
BECs bound by gravity in the laboratory.

The self-gravitating compact objects formed with BEC, called BEC stars, have attracted a lot of interest from the research community recently ~\cite{Jones_2001,Chavanis_2011,Chavanis_2011_paper2,Chavanis:2011cz,Chavanis:2014hba,Kling:2017mif,Kling:2017hjm,Annulli:2020ilw,Chavanis:2022fvh}. Gravitationally bounded BECs are also considered to be possible candidates of dark matter, which represents a significant amount of the total matter in the universe and several astrophysical and cosmological studies
in this direction have been carried out~\cite{Boehmer:2007um,Lee:2008jp,Harko:2011xw,Harko:2011jy,chavanis2012growth,Li:2012sg,Freitas:2012yr,Rindler-Daller:2013zxa,Guzman:2013rua,Madarassy:2014jfa,Eby:2015hsq}.
Bose condensates are also proposed to exist in the interior of neutron stars amidst fermions and
these possibilities are studied in detail~\cite{Glendenning:1997wn}. Neutron stars with BEC matter are viable because of the fact that neutrons in the star can exist in super-fluid phase, with the particles being treated as composite bosons through Cooper pair formation~\cite{Pethick:2015jma}.
In the present work, we concentrate on BEC stars for the analysis.

The dynamics of a self-gravitating BEC at zero temperature is described by the Gross-Pitaevskii (GP) equation~\cite{Gross1961,pitaevskii1961vortex} coupled with the Poisson equation in Newtonian approximation.
There are several studies on astrophysical implications of BEC using Newtonian analysis~\cite{Jones_2001,Chavanis_2011,Chavanis_2011_paper2,Chavanis:2011cz,Chavanis:2014hba,Kling:2017mif,Kling:2017hjm,Annulli:2020ilw}.
For example, in Ref.~\cite{Chavanis_2011},
the authors study the structure and stability of a self-gravitating BEC system with short-range interactions, and obtained an approximate analytical expression for the mass-radius relation.
These analytical relations were compared with the exact mass-radius relations obtained numerically by solving the equation of hydrostatic equilibrium and a good agreement between the two were found~\cite{Chavanis_2011_paper2}.
 The analyses conducted in Refs.~\cite{Chavanis_2011,Chavanis_2011_paper2} are motivated by the idea that dark matter could be a self-gravitating BEC. In Ref.~\cite{Boehmer:2007um}, dark matter is being described as a non-relativistic Bose-Einstein condensate gas with polytropic EoS.
There are also studies involving semi-relativistic formalisms, in which the stellar matter is described using a non-relativistic equation of state (EoS), while general relativistic approach is used to obtain the stellar structure configurations~\cite{Chavanis:2011cz,Chavanis:2014hba}.
\par
The possibility of existence of rotating BEC stars has also been considered in several works.
Rotating boson stars have been studied in Ref.~\cite{Silveira:1995dh} within the Newtonian approximation and their analysis suggests that the non-relativistic rotation can be applied only to smaller bosonic objects. Ref.~\cite{Wang:2001wq} studies cold Bose stars comprising of a dilute BEC using the non-relativistic GP equation (together with Poisson equation) and perturbative solutions of such slowly rotating stars were obtained. Slowly rotating BEC stars in the Newtonian limit obeying polytropic EoS have also been studied by solving the Lane-Emden equation~\cite{Chavanis:2011cz}.
The impact of slow rotation on the astrophysical properties of BEC dark matter halos were also determined using the non-relativistic approach~\cite{Zhang:2018okg}.
It was noted that the general relativistic effects impose strong constraints on the global parameters of the BEC stars and the obtained values within the non-relativistic formalism may exceed this stability limit~\cite{Chavanis_2011,Danila:2015qla}.
Relativistic BEC stars obeying the Colpi-Shapiro-Wasserman (CSW) EoS introduced in Ref.~\cite{Colpi:1986ye} were studied in Ref.~\cite{Chavanis:2011cz} and it was speculated that the observed massive neutron stars~\cite{Demorest:2010bx,Antoniadis:2013pzd} are composed of BECs.
Further, the analysis of Ref.~\cite{Chavanis:2011cz} was generalized to include spin and focused on studying the
observational constraints on the structural properties of spinning relativistic BEC stars obeying the CSW EoS ~\cite{Mukherjee:2014kqa} by employing the 
Rapidly Rotating Neutron Star (RNS) code~\cite{Stergioulas:1994ea},
based on general relativistic treatment.
The electromagnetic and thermodynamic properties of thin accretion disks around rotating pure BEC stars with a polytropic EoS have been investigated in Ref.~\cite{Danila:2015qla}, by using the RNS code to obtain the rotating configurations of the star.
In the present work, we intend to study the static and slowly rotating BEC stars by considering the general relativistic effects.

 Further, BEC has been generalized to finite temperatures and has been studied in the contexts of cosmological dark matter~\cite{Harko:2011dz,Harko:2012ay} and static compact objects~\cite{Latifah:2014ima,Gruber:2014mna,Angulo:2022gpj}, recently.
 There are attempts to study boson stars with inclusion of chemical potential and temperatures~\cite{Ingrosso_1988,Bilic:2000ef,Bhatt:2009gv,Matos:2011kn}. The theory of condensate dynamics at finite temperature has been discussed in Refs.~\cite{Griffin:1996,zaremba1999dynamics, griffin_nikuni_zaremba_2009}.
A finite temperature BEC has been considered in Ref.~\cite{Harko:2012ay}, which is composed of two fluids, the pure condensate and thermal fluctuations. An EoS at finite temperature  describing such a BEC system has been obtained analytically. Further, they apply this description of BEC to study the cosmological evolution of finite temperature dark matter filled Universe in a flat Friedmann-Robertson-Walker geometry.
 Pure BEC stars described by a thermodynamically consistent finite temperature EoS were considered and the global properties of a static BEC star were studied~\cite{Latifah:2014ima}. Recently, the astrophysical properties of relativistic BECs at finite temperature with the inclusion of magnetic field were also investigated~\cite{Angulo:2022gpj}.
 Further, Ref.~\cite{Gruber:2014mna} considers a BEC with repulsive contact and attractive gravitational interactions in the interior of neutron stars by the formation of Cooper pairs and macroscopic properties of the star have been evaluated at finite temperature.
 They describe the BEC stellar matter at the core to be composed of pure condensate together with a non-condensate cloud of excitations due to the presence of temperature.
 Analysis of pure BEC stellar equilibrium at finite temperature employing this EoS
 will be of interest.
  We note that, this finite temperature EoS is different from the one derived in Ref.~\cite{Latifah:2014ima} based on a thermodynamical method.
 Moreover, the study of macroscopic properties of rotating BEC stars at finite temperature have not been attempted before.
This sets the motivation for the present analysis.

 In the current work, we intend to analyse the effect of temperature on the global properties of static and rotating BEC stars.
 We employ the recently developed finite temperature BEC equation of state for the stellar matter, obtained using the generalised GP equation with repulsive contact and attractive gravitational interactions with approximations based on the semi-classical Hartree-Fock theory~\cite{Gruber:2014mna}.
The stellar matter now is considered to be composed of Bose condensed state together with a small fraction of excitations of BEC due to the presence of temperature.
In our analysis, we treat gravity within the framework of general relativity. The properties of static BEC stars are analysed for different temperatures by solving the Tolman-Oppenheimer-Volkoff equations~\cite{Tolman:1939jz,Oppenheimer:1939ne}. We include rotation in the analysis through Hartle-Thorne slow rotation approximation~\cite{Hartle:1967he,Hartle:1968si}, which is a well known perturbative approach and has been widely used  in studies of compact objects~\cite{Baubock:2013gna,Barausse:2015frm,Jyothilakshmi:2022hys}. To the best of our knowledge, this appropriate relativistic slow rotation approximation has not been employed so far to study BEC stars.
\par
The paper is organised as follows. In section~\ref{BEC_EOS}, we review the BEC equation of state at finite temperature used for the analysis. Then, we discuss the general relativistic stellar structure equations for static and slowly rotating stars in section~\ref{stellar_structure}. Next, we present the results in section~\ref{results}. Finally, we summarise our results and conclusions in the last section.

\section{Equation of state for finite temperature BEC\label{BEC_EOS}}

In this section, we briefly review the formalism to estimate the equation of state for a non-rotating BEC subjected to repulsive contact and attractive gravitational interactions at finite temperature as derived in Ref.~\cite{Gruber:2014mna}. The formalism to describe such a non-rotating BEC system is based on the studies in Refs.~\cite{Griffin:1996,zaremba1999dynamics,griffin_nikuni_zaremba_2009}.
At zero temperature, BEC is described by a macroscopic wave-function $\Psi(\textbf{r},t)$, whose evolution is given by the Gross-Pitaevskii (GP) equation~\cite{pitaevskii_2003,Pethick:2002}. A finite temperature BEC system consists of condensate particles together with a non-condensate cloud of thermal fluctuations~\cite{zaremba1999dynamics,Harko:2012ay,Gruber:2014mna}. The thermal cloud comprises of excitations from the condensate due to the presence of temperature and these excitations vanish with the decrease in temperature resulting in a system with pure condensate. The description of such a system begins by considering BEC at zero temperature and extending it to finite temperature.
 \par
The dynamics of a Bose-Einstein condensate is described by the Heisenberg equation of motion for the Bose field operator $\hat{\psi}(\mathbf{r},t)$ and is given by~\cite{zaremba1999dynamics,Pethick:2002,Griffin:1996}
\begin{align}\label{gpet}
    i \hbar \frac{\partial \hat{\psi}(\mathbf{r},t)}{\partial t}=\Bigg[&-\frac{\hbar^2}{2 m}{\nabla}^2+ \Phi(\mathbf{r},t)  \nonumber\\
    &+g\hat{\psi}^{\dag}(\mathbf{r},t) \hat{\psi}(\mathbf{r},t)\Bigg] \hat{\psi}(\mathbf{r},t),
\end{align}
where $\Phi(\mathbf{r},t)$ is the Newtonian gravitational potential.
The quantity $g = 4\pi a\hbar^2/m$ represents the strength of the repulsive contact interaction with $a$ being the $s$-wave scattering length of bosons in the system;  $m$ denotes the mass of condensate particles
and $G$ is the Newton's gravitational constant.
The Heisenberg equation is extended to finite temperature by considering the effects of non-condensate atoms, referred to as the thermal cloud, along with the condensate particles.
Now, by assuming Bose broken symmetry, the field operator can be decomposed as~\cite{Griffin:1996}
\begin{equation}
\hat{\psi}(\textbf{r},t)=\Psi(\textbf{r},t)+\hat{\Psi}_\text{th}(\textbf{r},t),
\end{equation}
where, the expectation value of the Bose field operator denotes the condensate wave-function $\Psi(\textbf{r},t) \equiv \langle\hat{\psi}(\textbf{r},t)\rangle$ and $\hat{\Psi}_\text{th}(\textbf{r},t)$ is the non-condensate field operator with $\langle \hat{\Psi}_\text{th}(\textbf{r},t)\rangle = 0$.
By taking the average of Eq.~(\ref{gpet}) with respect to a broken symmetry non-equilibrium ensemble, we obtain an exact equation of motion for the condensate wave-function $\Psi(\textbf{r},t)$ given by~\cite{Griffin:1996}
\begin{align}\label{gpett}
    i \hbar \frac{\partial \Psi(\textbf{r},t)}{\partial t} =&\left[-\frac{\hbar^2}{2 m} {\nabla}^2 + \Phi(\textbf{r},t)\right] \Psi( \textbf{r},t) \nonumber\\
    &+g\left\langle\hat{\psi}^{\dag}(\textbf{r},t) \hat{\psi}(\textbf{r},t) \hat{\psi}(\textbf{r},t)\right\rangle .
    \end{align}
Considering the expansion of the term,
\begin{eqnarray}\label{expansion}
    \hat{\psi}^{\dag}(\textbf{r},t) \hat{\psi}(\textbf{r},t) \hat{\psi}(\textbf{r},t)
     &=&|\Psi|^2 \Psi+2|\Psi|^2 \hat{\Psi}_\text{th} + \Psi^2 \hat{\Psi}^{\dag}_\text{th} \nonumber\\
    && + 2 \Psi \hat{\Psi}^{\dag}_\text{th} \hat{\Psi}_\text{th} + \Psi^{\dag} \hat{\Psi}_\text{th} \hat{\Psi}_\text{th}\nonumber\\
&&+{\hat{\Psi}^{\dag}_{\text{th}}} \hat{\Psi}_\text{th} \hat{\Psi}_\text{th},
\end{eqnarray}
we proceed to find its expectation value by noting the number densities of the condensate and the thermal cloud respectively as
 \begin{eqnarray}\label{n}
     n(\textbf{r},t)&=&\left\langle{\Psi}^{\dag}(\textbf{r},t) {\Psi}(\textbf{r},t)\right\rangle,\\
     n_\text{th}(\textbf{r},t)&=&\left\langle \hat{\Psi}^{\dag}_\text{th}(\textbf{r},t) {\hat{\Psi}_\text{th}}(\textbf{r},t)\right\rangle.
 \end{eqnarray}
The second and third terms on the right-hand side of Eq.~\eqref{expansion} vanish, since the average value of thermal fluctuations is zero because of the assumed broken symmetry. Further, by denoting the mass densities of condensate and non-condensate parts as $\rho(\textbf{r},t)=m n(\textbf{r},t)$ and  ${\rho}_\text{th}(\textbf{r},t)=m{n}_\text{th}(\textbf{r},t)$ respectively, and the off-diagonal (anomalous) mass density as $\rho_{a}(\textbf{r},t)=m n_a(\textbf{r},t)=m\langle{\hat{\Psi}_\text{th}}(\textbf{r},t) {\hat{\Psi}_\text{th}}(\textbf{r},t)\rangle$, we get the following expression:
\begin{align}  \label{threepsi}
    \left\langle\hat{\psi}^{\dag}(\textbf{r},t) \hat{\psi}(\textbf{r},t) \hat{\psi}(\textbf{r},t)\right\rangle =& \frac{1}{m} \rho\Psi+2 \frac{1}{m} {\rho}_\text{th}\Psi \\
    &+\frac{1}{m}\rho_{{a}} \Psi^{\dag}+\left\langle\hat{\Psi}^{\dag}_\text{th} \hat{\Psi}_\text{th} {\hat{\Psi}_\text{th}}\right\rangle. \nonumber
\end{align}
Substituting Eq.~(\ref{threepsi}) in Eq.~(\ref{gpett}) yields the equation of motion for $\Psi$ - the \textit{generalised} GP equation~\cite{zaremba1999dynamics} :
    \begin{align}\label{eqn of motion}
        i \hbar \frac{\partial \Psi(\textbf{r},t)}{\partial t} &= \Bigg[-\frac{\hbar^2}{2 m} {\nabla}^2 + \Phi(\textbf{r},t)+gn(\textbf{r},t) \nonumber\\
        &+2 g{n}_\text{th}(\textbf{r},t)\Bigg] \Psi(\textbf{r},t)
         +g{n_a}(\textbf{r},t) \Psi^{\dag}(\textbf{r},t) \nonumber \\
         &+g\left\langle\hat{\Psi}^{\dag}_\text{th}(\textbf{r},t) \hat{\Psi}_\text{th}(\textbf{r},t)\hat{\Psi}_\text{th}(\textbf{r},t)\right\rangle.
    \end{align}
We now assume a Madelung representation of the condensate wave function with the phase term $S(\mathbf{r},t)$ having dimension of action~\cite{Pethick:2002},
\begin{equation}\label{madelung}
    \Psi(\textbf{r},t)=\sqrt{n(\textbf{r},t)}\, e^{(i / \hbar)S(\textbf{r},t)}.
\end{equation}
Here, the phase factor is related to the velocity of the condensate as $\mathbf{v}(\mathbf{r},t)=\nabla S/m$. Substitution of the above form of the wave function in generalised GP equation, Eq.~(\ref{eqn of motion}), results in two equations corresponding to real and imaginary part as~\cite{zaremba1999dynamics},
\begin{align}
\label{continuity}
\frac{\partial n}{\partial t}+\pmb{\nabla} \cdot\left(n\mathbf{v}\right)=& \,\frac{2g}{\hbar} \operatorname{Im}\left[\left(\Psi^{\dag}\right)^2n_a  + \Psi^{\dag}\left\langle \hat{\Psi}^{\dag}_\text{th} \hat{\Psi}_\text{th} \hat{\Psi}_\text{th}\right\rangle\right] \nonumber \\
\frac{\partial S}{\partial t}=&\frac{{\hbar}^2}{2m\sqrt{n}}{\nabla}^2\sqrt{n}  - \Phi(\mathbf{r},t) - gn(\mathbf{r},t)\nonumber \\
 & - 2g{n_\text{th}}(\mathbf{r},t) - \frac{g}{n} \text{Re}\Big[(\Psi^{\dag})^2n_a \nonumber \\
 & +\Psi^{\dag}\left\langle \hat{\Psi}^{\dag}_\text{th} \hat{\Psi}_\text{th} \hat{\Psi}_\text{th}\right\rangle\Big] -\frac{1}{2} m{v}^2.
\end{align}

Now, using the several approximation schemes, as discussed in Ref.~\cite{Gruber:2014mna} and references therein, we simplify the above obtained equations.
We can utilize the Hartree-Fock-Bogoliubov (HFB) approximation to ignore the three-field correlation function $\left\langle \hat{\Psi}^{\dag}_\text{th} \hat{\Psi}_\text{th} \hat{\Psi}_\text{th}\right\rangle$. The dynamic Popov approximation can be used to ignore both $\left\langle \hat{\Psi}^{\dag}_\text{th} \hat{\Psi}_\text{th} \hat{\Psi}_\text{th}\right\rangle$ and anomalous mass density $\rho_{a}(\textbf{r},t)$. With the static Popov approximation, the fluctuations of the density of the thermal cloud are ignored by assuming that the non-condensate is always in static thermal equilibrium, so that ${n}_{\text{th}}(\textbf{r},t) \simeq{n}_{\text{th}}(\textbf{r}) $. Further, the time dependence in all other terms are neglected~\cite{Gruber:2014mna}.
Now, using these approximations,
Eqs.~\eqref{continuity} take the form of hydrodynamic continuity and Euler equations~\cite{zaremba1999dynamics}:
\begin{eqnarray}
    \frac{\partial n}{\partial t} + \pmb{\nabla}\cdot (n\mathbf{v}) &=& 0,\\
      mn\left[\frac{\partial\mathbf{v}}{\partial t}  + \pmb{\nabla}\left(\frac{v^2}{2}\right) \right]
      &=&  \frac{{\hbar}^2}{2m\sqrt{n}}{\nabla}^2\sqrt{n} - n\pmb{\nabla}\Phi(\textbf{r}) \nonumber\\
      &&  - gn\pmb{\nabla}\left(n + 2n_{\text{th}}\right),
    \label{eul_thermal}
\end{eqnarray}
respectively.
Further, we adopt the Thomas-Fermi approximation~\cite{Harko:2011dz}, which neglects the kinetic energy term of the condensate $i.e.,\,-\hbar^2\nabla^2/2m$.
Therefore, the quantum correction stress tensor term proportional to ${\nabla}^2\sqrt{n}$ in Eq.~(\ref{eul_thermal}) also gets neglected.
Eq.~(\ref{eul_thermal}) is compared with the general Euler equation to obtain the gradient of condensate pressure as~\cite{Gruber:2014mna}
\begin{equation} \label{pressure}
    \pmb{\nabla} p = g n \pmb{\nabla} (n + 2 n_{\text{th}}).
\end{equation}
The expression for non-condensate density can be obtained by integrating the Bose-Einstein distribution over the momentum space in spherical polar coordinates $(r_k,\theta_k,\phi_k)$,
\begin{equation} \label{n_thermal}
n_{\text{th}}(r_k)=\int \frac{d^3k}{(2\pi)^3}\frac{1}{e^{\beta[ \epsilon_{{k}}(r_k) - \mu] }-1},
\end{equation}
where, $\epsilon_{{k}}(r_k)$ represents the energy of the thermal excitations and $\mu$ denotes the chemical potential of the condensate. Here, $\beta =(k_BT)^{-1}$, with $T$ as the temperature of the system and $k_B$ as the Boltzmann constant.
Now, to evaluate Eq.~\eqref{n_thermal}, we employ the semi-classical Hartree-Fock equations of motion for a system with contact and gravitational interactions~\cite{Gruber:2014mna},
\begin{eqnarray}
 \epsilon_{{k}}({r_k}) &=& \frac{{\hbar}^2k^2}{2m} + 2g[n({r_k}) + n_{\text{th}}({r_k})] + \Phi(r_k), \\
     \mu &=& g[n({r_k}) + 2n_{\text{th}}({r_k})] +  \Phi(r_k),
\end{eqnarray}
where, $\Phi(r_k)$ is now given by
\begin{equation}
    \Phi({r_k})= - \int d^3r_k'\frac{Gm^2}{|{r_k}-{r_k}'|} \left[ n(r'_k)+n_{th}(r'_k)\right].
\end{equation}
Using the above relations in the expression for thermal density and solving with the help of standard integrals, we get
\begin{equation} \label{nth}
   n_{\text{th}}(r_k) = \frac{1}{\lambda^3}\,\zeta_{3/2}\left[e^{-\beta g n(r_k)}\right];
\end{equation}
where, $\lambda =  \sqrt{(2\pi\beta{\hbar}^2)/m}$ is the thermal de Broglie wavelength and $\zeta$ represents the polylogarithmic function
\begin{equation}
    \zeta_{\nu}[z] = \sum_{n=1}^{\infty}\frac{z^n}{n^{\nu}},
\end{equation}
with index $\nu$.
Substituting Eq.~\eqref{nth}, we integrate out Eq.~\eqref{pressure} to obtain the pressure of the condensate including the effects of the thermal cloud~\cite{Gruber:2014mna}:
\begin{align} \label{eos}
        p =&\frac{g\rho^2}{2m^2} + \frac{2g\rho}{m\lambda^3}\,\zeta_{3/2}\left[e^{-\beta g\rho/m}\right] \nonumber \\
        &+\frac{2}{\beta \lambda^3} \zeta_{5/2} \left[e^{-\beta g\rho/m}\right] -\frac{2}{\beta\lambda^3}\,\zeta_{5/2}\left[ 1\right].
\end{align}
Here, the first term denotes the pressure of the pure condensate in the absence of thermal fluctuations. The second and third terms represent the contribution of thermal excitations to the pressure of the condensate. Further, the constant term ensures that the condensate pressure vanishes in the limit $\rho \rightarrow 0$.

\begin{figure}[t!]
\centering
\includegraphics[width=\linewidth]{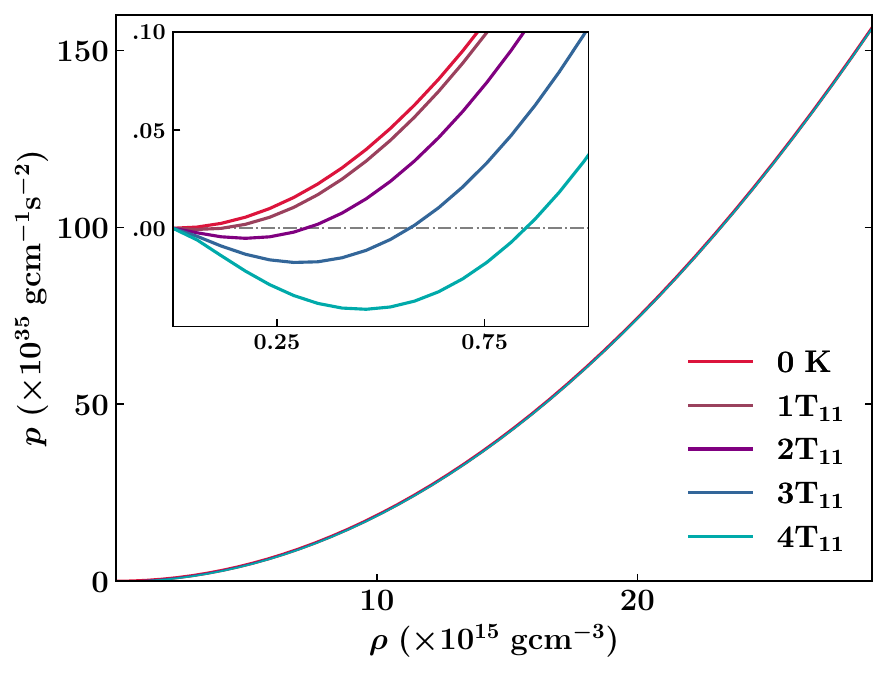}
\caption{BEC equation of state with the effect of thermal fluctuations for different temperatures. Inset highlights the effect of temperature in lower densities.
}
\label{Fig:EOS}
\end{figure}
We plot the above obtained BEC EoS for different temperatures in Fig.~\ref{Fig:EOS}. Here, we take the mass of the condensate particle to be $m=2m_n$,
where $m_n$ is the mass of a nucleon,
considering the possibility of two of them forming an equivalent Cooper pair and act as a boson \cite{Chavanis:2011cz}. The scattering length is taken to be $a=1$ fm \cite{Gruber:2014mna}. Also, following Ref.~\cite{Gruber:2014mna}, in our studies, we focus on relevant temperatures between $(1-4) \times 10^{11}\hspace{0.1cm}$K.
\par
In Fig.~\ref{Fig:EOS}, the curve plotted for $T=0$ K represents the pressure of pure condensate. Note that this corresponds to the polytropic EoS used in Ref.~\cite{Chavanis:2011cz}.
We observe that the presence of temperature in the system decreases the condensate pressure. This is due to the fact that temperature in the system results in thermal excitations of few fraction of condensate particles.
 It can also be seen that the pressure becomes negative for small densities. This is a consequence of the Thomas-Fermi approximation, which neglected the quantum pressure term of the condensate in the calculations.
 This term becomes important at the edge of the star, where the density of the condensate is low. Addition of this quantum pressure term would presumably correct the unphysical negative pressure observed for small densities~\cite{Gruber:2014mna}. Moreover, we note that a zero-temperature treatment would be sufficient for temperatures below $10^{11}\hspace{0.1cm}$K, since the thermal fluctuations are negligible within that temperature range. Futhermore, Fig.~\ref{Fig:EOS} shows that although the presence of temperature decreases the pressure of the condensate for a fixed density, this decrement is observed to be very small implying small deviation of the pressure from the $T=0$ K case for the range of $\rho$ values considered.
For example, at $\rho = 2\times 10^{16}$ g/cm$^3$, the deviation of the condensate pressure from the zero temperature case is $\sim 0.27\%$ when $T=4\times 10^{11}$ K; while the deviation is only $\sim 0.008\%$ for $T=1\times 10^{11}$ K.
The zero temperature polytropic EoS has been employed to study rotating BEC stars within Newtonian~\cite{Boehmer:2007um} and general relativistic~\cite{Danila:2015qla} treatments.
In the present work,
we introduce the effect of slow rotation to the finite temperature BEC system described by Eq.~\eqref{eos} via general relativistic treatment.
\par
We now proceed to calculate the stellar configuration dynamic equations within general theory of relativity to utilise the above obtained temperature dependent EoS to study BEC star profiles.

\section{Stellar structure equations \label{stellar_structure}}

 We employ a general relativistic treatment to study the stellar structure equations. We first calculate the non-rotating configurations of the star and then use them to obtain the slowly rotating configurations.  In this section,  we follow the metric convention $g_{\mu\nu}=\textrm{diag}(-1,1,1,1)$ and use velocity of light $c=1$. The metric for a spherically symmetric static relativistic star can be expressed as
 \begin{align}
   ds^{2} &= -e^{\nu} dt^{2}+e^{ \Lambda} dr^{2}+r^{2}\left(d\theta ^{2}+\sin ^{2}\theta d\phi ^{2}\right).
\end{align}
Here, $\nu(r)$ and $\Lambda(r)$ are the metric functions.
Assuming the stellar matter to be described by perfect fluid with energy density $\rho$ and pressure $p$, the Einstein field equations result in the Tolman-Oppenheimer-Volkoff (TOV) equations, which are given as \cite{Tolman:1939jz,Oppenheimer:1939ne}
\begin{subequations}
  \begin{align}\label{TOV-1}
    \frac{dp}{dr} &=-\frac{G(\rho+p)(M+4\pi r^3 p)}{r^2(1-2GM/r)},\\
    \label{TOV-2}
    \frac{dM}{dr} &= 4\pi r^2 \rho.
  \end{align}
\end{subequations}
The above coupled differential equations are solved from the centre to the surface of the star by providing an equation of state for the stellar matter. We employ the EoS given by Eq.~\eqref{eos} with the assumption that the condensate can be described by a non-relativistic EoS. The pressure and mass at the centre of the star are $p_c = p(\rho_c)$ and $M_c = 0$ respectively. The pressure $p$ vanishes as it approaches the surface ($r=R$) of the star. The mass of the star is then obtained as $M(r=R)$. By varying the central density $\rho_c$, we can get the maximum mass (radius) stellar configuration possible for the given EoS.
\par
The solutions $p(r)$ and $M(r)$ obtained for a given central density are then used to solve for a slowly rotating BEC star.  For this we make use of the Hartle Thorne approximation \cite{Hartle:1967he,Hartle:1968si} in which rotation is treated as a small perturbation on the metric of the non-rotating star:
\begin{align}\label{hmetric}
    ds^2 = &-e^\nu[1+2(h_0+h_2P_2)]dt^2\nonumber\\
     &+ \frac{1+2G(m_0+m_2P_2)(r-2GM)^{-1}}{1-2GM/r}dr^2 \nonumber\\
    &+ r^2\left[1+2(v_2-h_2)P_2\right]\left[d\theta^2+\sin^2\theta(d\phi-\omega dt)^2\right]\nonumber\\
   & + O(\Omega^3).
\end{align}
Here, $P_2=P_2(\cos\theta)$ is the second order Legendre polynomial, $\omega$ is the frame dragging frequency which is proportional to $\Omega$ and is a function of $r$, while $h_0,m_0,h_2,m_2,p_2$ and $v_2$ are functions of $r$ that are proportional to $\Omega^2$.
The angular velocity relative to the local inertial frame, $\bar{\omega}$ ($ = \Omega - \omega$ ) is obtained by solving the second order differential equation
\begin{equation}\label{omega}
    \frac{1}{r^4}\frac{d}{dr}\left(r^4j\frac{d\Bar{\omega}}{dr}\right) +\frac{4}{r}\frac{dj}{dr}\bar{\omega}=0,
\end{equation}
where,
\begin{equation}
    j =e^{-\nu/2}
    {\left(1-2GM/r\right)}^{1/2}.
\end{equation}
Eq.~(\ref{omega}) is integrated from the centre to the surface of the star with the boundary conditions: $\bar{\omega}=\omega_c$ and $d\Bar{\omega}/dr = 0$.
The angular momentum $J$ and the angular velocity  $\Omega$ corresponding
to $\omega_c$ are
\begin{equation}
    J=\frac{1}{6}R^4\left(\frac{d\Bar{\omega}}{dr}\right)_{r=R}, \qquad \Omega= \Bar{\omega}(R) + \frac{2J}{R^3}.
\end{equation}
In order to obtain a different value of angular velocity, the function $\bar{\omega}(r)$ is re-scaled as
\begin{equation}
    {\Bar{\omega}(r)}_{new} ={\Bar{\omega}(r)}_{old}\left (\frac{\Omega_{new}}{\Omega_{old}} \right).
\end{equation}
The deformation of stellar structure as a result of rotation can be obtained in terms of $\xi(r,\theta)$~\cite{Hartle:1967he,Hartle:1968si}:
\begin{equation}
    \xi(r,\theta) = \xi_0(r) + \xi_2(r) P_2.
\end{equation}
Here, the first term corresponds to the spherical ($l=0$) deformation and the second term corresponds to the quadrupole ($l=2$) deformation. In our study, we focus only on the spherical deformation $\xi_0(r)$, which can be obtained by solving the mass perturbation factor ($m_0$) and the pressure perturbation factor ($p_0^*$) equations given as~\cite{Hartle:1967he,Hartle:1968si}
\begin{subequations}
    \begin{align} \label{HT_m}
        \frac{dm_0}{dr} =& 4\pi r^2\frac{d\rho}{dp}(\rho+p)p_0^* + \frac{1}{12G}j^2 r^4 \left(\frac{d\Bar{\omega}}{dr}\right)^2  \nonumber\\
        &- \frac{1}{3G}r^3 \frac{dj^2}{dr}\Bar{\omega}^2,\\
        \frac{dp_0^*}{dr} =& - \label{HT_p}\frac{Gm_0(1+8\pi Gr^2p)}{r^2{(1-2GM/r)}^2} - \frac{4\pi G (\rho+p)r}{(1-2GM/r)}p_0^*\nonumber \\
        &+ \frac{1}{12}\frac{r^3j^2}{(1-2GM/r)}{\left (\frac{d\Bar{\omega}}{dr}\right)}^2 + \frac{1}{3}\frac{d}{dr}\left(\frac{r^2 j^2 {\Bar{\omega}^2}}{1-2GM/r}\right).
    \end{align}
\end{subequations}

The above differential equations are to be integrated from the centre to the surface of the star, with the boundary conditions that at the centre of the star $m_0=p_0^* =0$. The density profile of the slowly rotating star is then obtained as
\begin{equation}\label{density profile}
     \rho_{rot}(r) = \rho_{stat}(r) - \frac{d\rho_{stat}(r)}{dr} \xi_0(r),
\end{equation}
where,
\begin{equation}
    \xi_0(r) = -p_0^*(r)\frac{\rho(r)+p(r)}{dp(r)/dr}.
\end{equation}
The gravitational mass ($M^*$) and the radius ($R^*$) of the slowly rotating star are given by
\begin{subequations}
\begin{align}
     M^* &= M(R) + m_0(R) + \frac{J^2}{R^3},  \label{24a}\\
   R^* &= R +  \xi_0(r)\label{24b}.
\end{align}
\end{subequations}
Here, $M(R)$ and $R$ are the mass and radius of a static star obtained from TOV equations for a given central density.
\par
Once, we prescribe the equation of state, by solving the above obtained set of equations for static and rotating cases, the corresponding stellar structure configurations can be obtained.

\section{Results and Discussions\label{results}}

\begin{figure}[t]
\centering
\includegraphics[width=\linewidth]{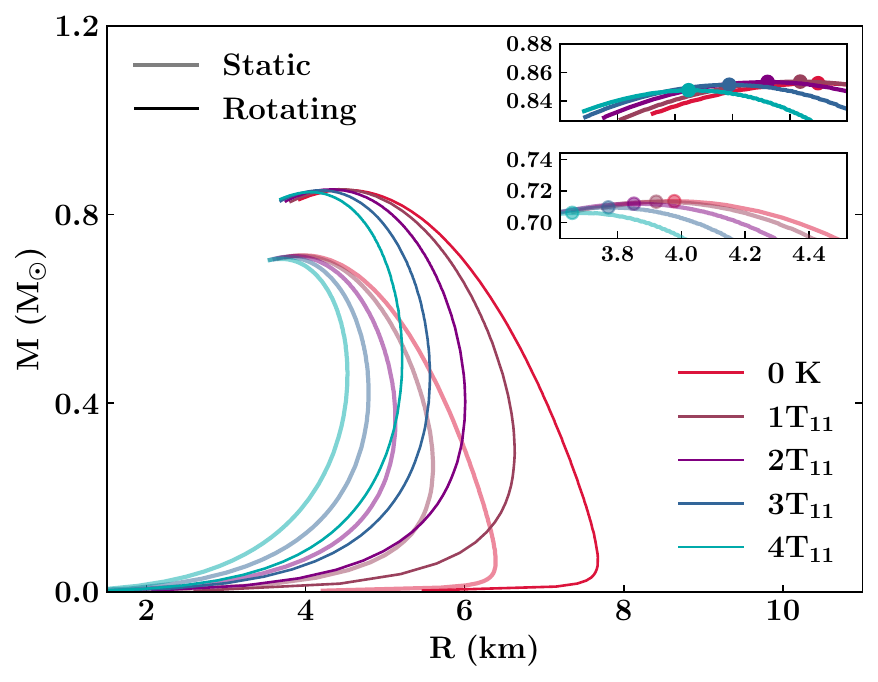}
\DeclareGraphicsExtensions{.pdf,.jpg}
\caption{Mass of the BEC star as a function of radius for different temperatures ($T_{11}=10^{11}$ K). The dashed curves correspond to the static stellar configurations. Solid curves correspond to the BEC star configurations rotating slowly with their Keplerian angular velocity $\Omega_K$. Inset shows the maximum masses in each case.}
\label{Fig:MR_plot}
\end{figure}
\begin{figure*}[t]
 \centering
 \subfigure[]{\includegraphics[width=0.45\linewidth]{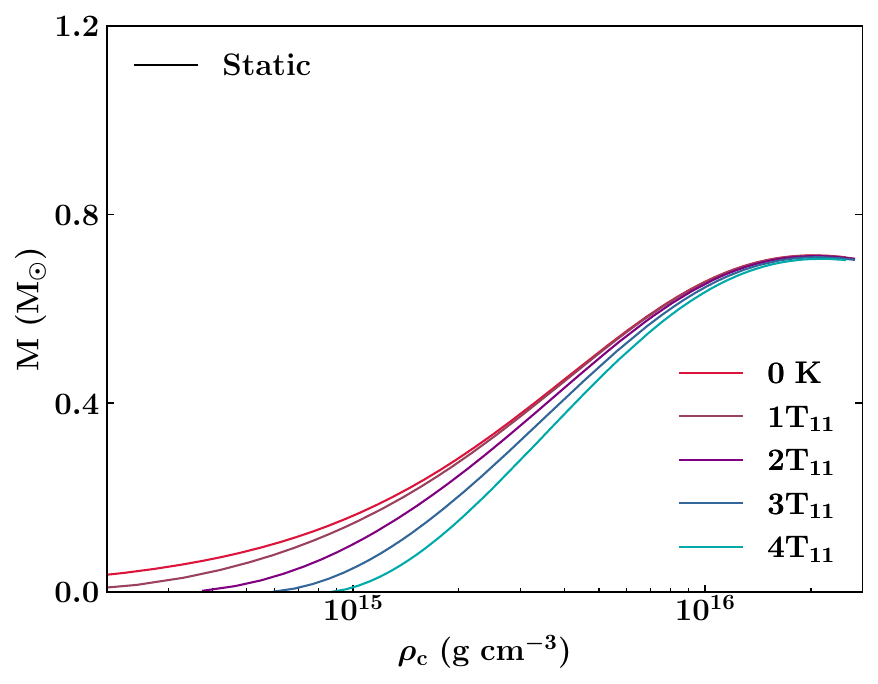} \label{rhoc_m_static}}\qquad
  \subfigure[]{\includegraphics[width=0.45\linewidth]{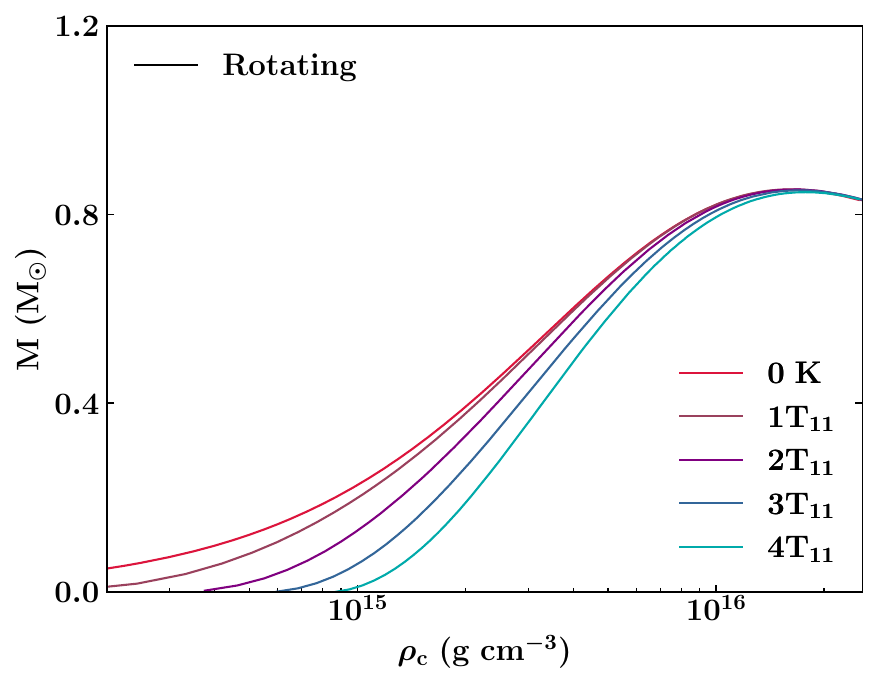}\label{rhoc_m_rot}}
  \caption{Mass of the (a) static and (b) slowly rotating stellar configurations as a function of central density for different temperatures ($T_{11}=10^{11}$ K).}
\end{figure*}
We numerically integrate the coupled stellar structure differential equations given by Eqs.~\eqref{TOV-1},~\eqref{TOV-2} and Eqs.~\eqref{HT_m},~\eqref{HT_p} using the temperature dependent BEC EoS, Eq.~\eqref{eos}, to obtain the static and rotating stellar configurations respectively. For brevity, we have used $M$ and $R$ to denote the maximum mass and corresponding radius in both static and rotating cases.
We are ignoring the possibility of formation of vortices in this analysis~\cite{Zhang:2018okg,Danila:2015qla}.
 We study the global properties of these configurations for different temperature values as taken in section~\ref{BEC_EOS}, ranging from $1T_{11}-4T_{11}$, where $T_{11} = 10^{11}$ K. As noted before \cite{Gruber:2014mna}, a zero temperature treatment would be sufficient for temperatures below $10^{11}$ K as the effect of temperature on EoS is negligible in that range.
Initially we keep $m=2m_n$, where $m_n =1.675\times 10^{-24}$g is the mass of a nucleon and scattering length as $a=1$ fm in our analysis.

In Fig.~\ref{Fig:MR_plot}, we plot the mass of both the static and slowly rotating BEC stars as a function of radius corresponding to different temperatures. For the static case, at $T=0$ K, we obtain a maximum mass $M = 0.71M_{\odot}$, with $M_{\odot}$ being the mass of the Sun, with radius $R= 3.97$ km corresponding to the central density $\rho_c = 2.03 \times 10^{16}$ g/cm$^3$.
Our results are in agreement with the zero temperature static general relativistic BEC stars studied in Ref.~\cite{Chavanis:2011cz}.
While coming to the finite temperature case, we see that static stellar equilibria are achieved at reduced radii and masses; though form of the mass-radius curve does not change appreciably. For example,
the radii corresponding to a $0.65 M_{\odot}$ star are $4.71$ km, $4.56$ km, $4.40$ km and $4.24$ km respectively for the temperatures $T=(1,\,2,\,3,\,4)T_{11}$.
This is due to the fact that increase in temperature results in a lower value of pressure for a given energy density (see inset of Fig.~\ref{Fig:EOS}), implying a softer equation of state. Softening of the EoS is a consequence of the presence of thermal fluctuations in the star. A softer EoS is known to result in lower maximum mass-radius models \cite{Glendenning:1997wn}.
Interestingly, the maximum mass and corresponding radius of the static BEC star does not vary significantly for the different temperatures considered in our analysis from the zero temperature case; as seen from the inset of the figure.
The maximum mass values corresponding to the temperatures $(1,2,3,4)T_{11}$ are $0.71 M_{\odot}$, $0.71 M_{\odot}$, $0.71 M_{\odot}$ and $0.70 M_{\odot}$ respectively.
We note that this particular behaviour was observed in the study conducted on static BEC stars at finite temperature employing a different EoS in Ref. \cite{Latifah:2014ima}.

\begin{figure*}[t]
  \centering
  \subfigure[]{\includegraphics[width=0.45\linewidth]{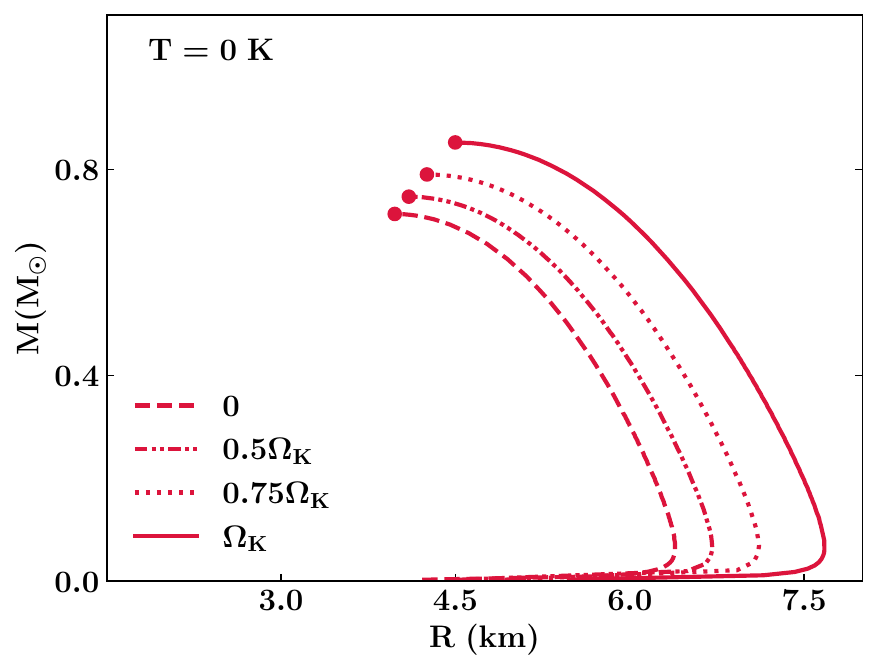} \label{MR_0K_omega}}\qquad
  \subfigure[]{\includegraphics[width=0.45\linewidth]{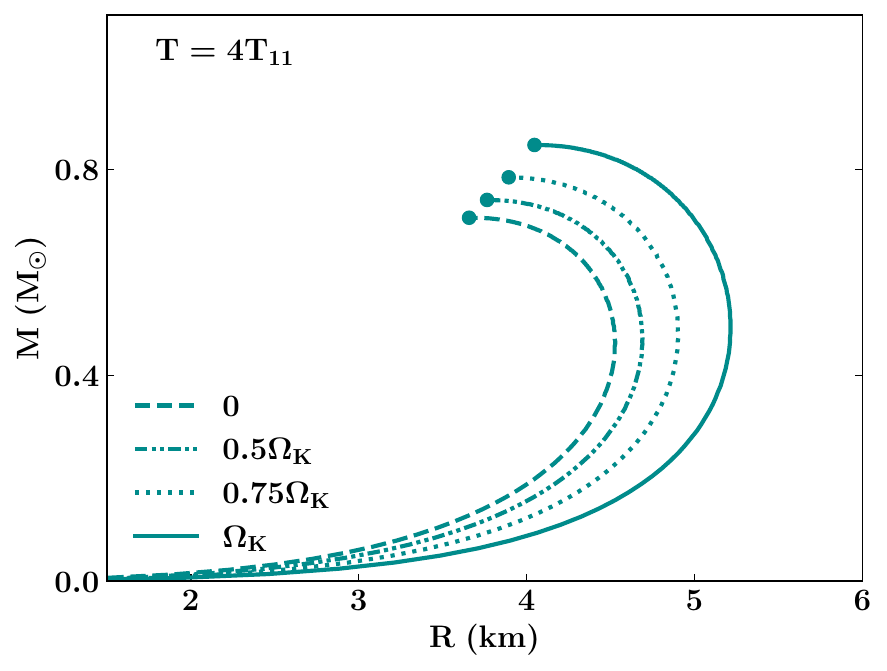}\label{MR_4K_omega}}
  \caption{Mass of the slowly rotating BEC star as a function of radius by varying the angular velocities, for (a) $T=0$ K and (b) $T=4T_{11}=4\times 10^{11}$ K.  }
  \label{Fig:Fig 9}
\end{figure*}
\par
Further, in Fig.~\ref{Fig:MR_plot}, we have expanded our analysis to incorporate rotation into the system. We allow the stellar equilibria to rotate in their Keplerian angular velocity $\Omega_K=\sqrt{GM/R^3}$; where $M$ and $R$ denotes the maximum mass and maximum radius of the static configurations.
Incorporating rotation into a system is likely to result in an additional centrifugal force. As a consequence of this, in order to maintain gravitational equilibrium, an increase in the maximum mass is anticipated and it is numerically obtained as $M = 0.85M_{\odot}$ for the zero temperature case.
This value corresponds to the central density $\rho_c = 1.6 \times 10^{16}$ g/cm$^3$ and radius $R= 4.47$ km.
Mere rotation increases the mass-radius profiles and maximum mass and radius of the star compared to the static case. However, as seen in the static case, increasing the system temperature results in reduction of mass-radius curves (due to softening of EoS) in rotational case too.
The radii corresponding to a $0.8M_{\odot}$ star are $5.28$ km, $5.08$ km, $4.87$ km and $4.66$ km respectively for the temperatures $T=(1,\,2,\,3,\,4)T_{11}$. In the rotating case too, we report that the maximum mass and radius for the stellar equilibria are not much altered by the presence of temperature considered. As seen from the inset  of Fig.~\ref{Fig:MR_plot}, the maximum masses corresponding to the temperature considered  $(1,2,3,4)T_{11}$ are  $0.85 M_{\odot}$, $0.84 M_{\odot}$, $0.84 M_{\odot}$ and $0.84 M_{\odot}$ respectively.
We note that $\Omega_K$  corresponding to the BEC stars with temperature $T=(0,1,2,3,4)T_{11}$ are {$(3.89,3.96, 4.07, 4.19, 4.37) \times 10^4\,\text{s}^{-1} $ }respectively for their  maximum mass configurations.\\

Next, we plot the variation of mass with central density for the static and rotating BEC stars in Fig.~\ref{rhoc_m_static} and \ref{rhoc_m_rot} respectively.
For both static and rotating case, the mass values obtained for the temperatures do not vary significantly at higher densities, whereas in the lower density region a significant difference is seen. The masses of rotating stars corresponding to $T=0$ K and $T=4 \times 10^{11}\hspace{0.1cm}$ K are obtained as $M = 0.39M_{\odot}$ and $M = 0.18M_{\odot}$ respectively, for $\rho_c = 2\times 10^{15}$ g/cm$^3$.
While for the static case, the masses corresponding to $T=0$ K and $T=4 \times 10^{11}\hspace{0.1cm}$ K are obtained as $M = 0.28M_{\odot}$ and $M = 0.15M_{\odot}$ respectively,
with the same value of central density. The maximum mass value configurations for $T=0$ K in static and rotating cases correspond to central densities $2.03 \times 10^{16}$ g/cm$^3$ and $1.6 \times 10^{16}$ g/cm$^3$ respectively. Introduction of temperature results in increase in the central density value corresponding to maximum masses. In the static case, the central density values corresponding to the maximum mass for the temperatures $(1,2,3,4)\times 10^{11}$ K are $(2.05,2.06,2.09,2.20) \times 10^{16}$ g/cm$^3$ respectively. And for the rotating case the maximum mass for the temperatures $(1,2,3,4)\times 10^{11}$ K corresponds to the central densities $(1.61,1.65,1.71,1.79) \times 10^{16}$ g/cm$^3$ respectively.
We also note that, with the increment in central densities, mass of the configurations also increases, while the corresponding radii decrease.

As we study the relationship between angular velocity and mass in rotating systems, we notice that the difference between static star and rotational mass grows with increasing angular velocity as illustrated in Fig.~\ref{Fig:Fig 9}.
This can be attributed to the fact that the increase in mass caused by rotation is directly proportional to the angular velocity. As the angular velocity increases, it results in a corresponding increase in the centrifugal force experienced within the system. To counterbalance this augmented centrifugal force, a larger mass is required.
We can read out from Fig. \ref{Fig:Fig 9} that as we increase the angular velocity from zero to $0.5\Omega_K$ and $0.75\Omega_K$, the maximum mass correspondingly increases to $M = 0.75 M_{\odot}$ and $M = 0.79 M_{\odot}$ respectively.
The radius values obtained for the same are $R= 4.09$ km and $R=4.25$ km respectively.
A similar trend is observed for finite temperatures.
As the star starts rotating and when the angular velocities increase from $\Omega = 0.5 \Omega_K$ to $\Omega = 0.75\Omega_K$, 
the maximum mass and corresponding radius values increase from $M = 0.74 M_{\odot}$ to $M = 0.78 M_{\odot}$ and $R=3.76$ km to $R=3.89$ km respectively, for $T=4T_{11}\hspace{0.1cm}$.
We also note that due to the presence of temperature the star attains a larger mass at lower densities compared to the zero temperature case for all values of angular velocities considered.
It must be highlighted that the relativistic treatment is quite important while considering the global properties of BEC stellar models. The mass of the static BEC star obtained by solving the Newtonian Lane-Emden equation for $\rho_c = 4 \times 10^{15}$ g/cm$^3$ is $0.74M_{\odot}$ and corresponding slow rotation (with $\Omega_K = 1.84\times 10^4 \,s^{-1}$) gives the value $1.11M_{\odot}$ \cite{Chandrasekhar1933}. Whereas, within our general relativistic treatment, correspondingly, we get the mass values as $0.45M_{\odot}$ and  $0.60M_{\odot}$ for static and slow rotation, respectively.
\par
\begin{figure}[t!]
\centering
\includegraphics[width=0.9\linewidth]{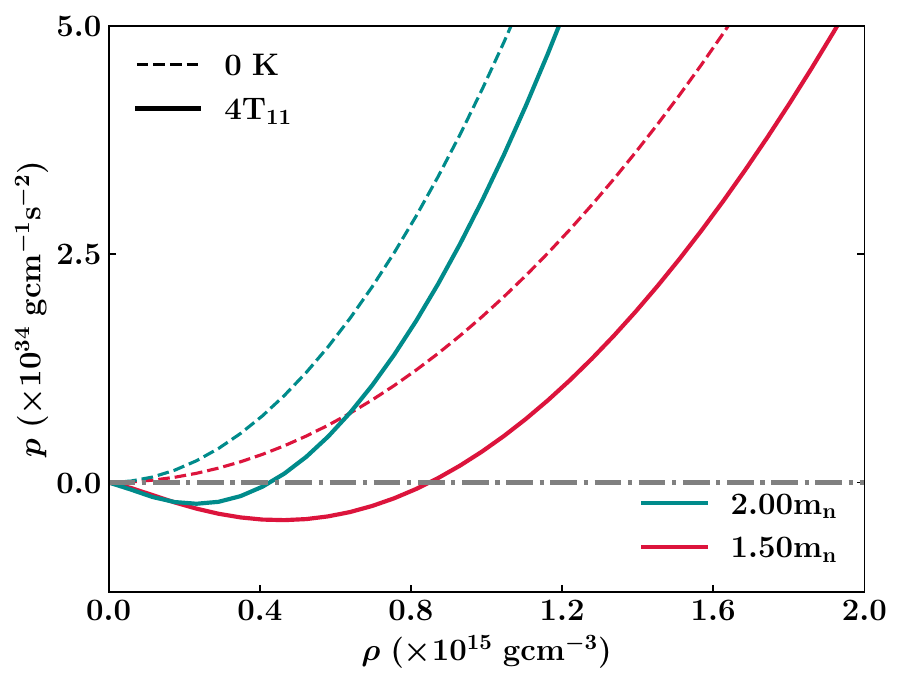}
\caption{BEC equation of state with different values of mass $m$ for $T=0$ K and $T=4\times 10^{11}$ K.}
\label{Fig:EOS_mn}
\end{figure}
\begin{figure*}[t]
  \centering
  \subfigure[]{\includegraphics[width=0.45\linewidth]{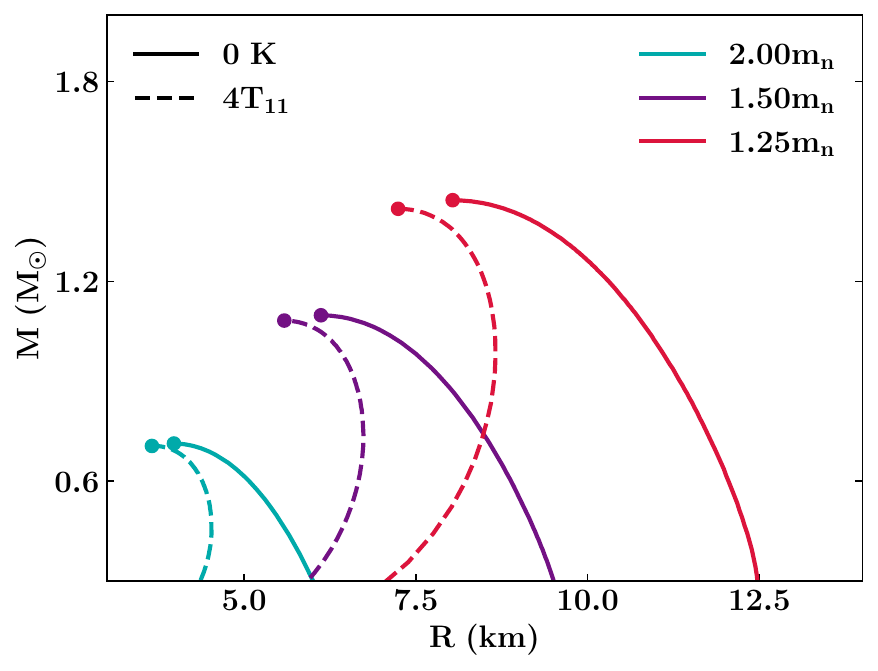}}
  \subfigure[]{\includegraphics[width=0.45\linewidth]{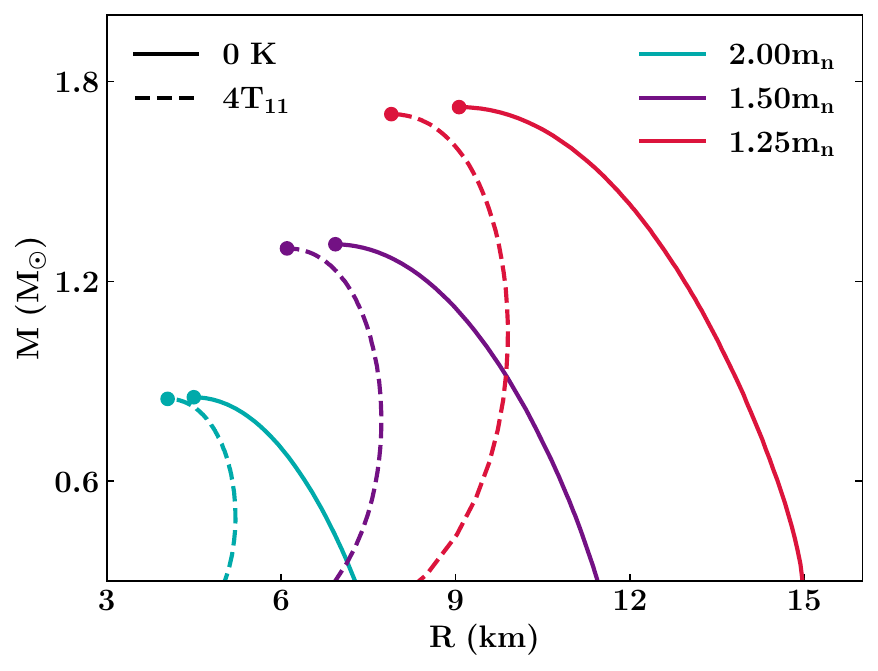}
  }
  \label{mn_4K}
  \caption{Mass-radius relations of (a) static and (b) rotating BEC stars for different values of the condensate mass $m$ (in terms of the nucleon mass $m_n$) at zero and finite temperature. Scattering length is kept as $a = 1$ fm.}
  \label{Fig:varied_mn}
\end{figure*}
Further, we study the effect of condensate mass $m$ on the static and rotating stellar configurations at finite temperature. First, we plot the BEC EoS for different values of boson mass: $m=2m_n$ and $m=1.5m_n$, with zero and finite temperatures in Fig. \ref{Fig:EOS_mn}. We can see that, lower the value of $m$, pressure increases for a given value of density resulting in a stiffer EoS. With $T=4T_{11}$ and for $m=1.5m_n$, EoS remains stiffer even compared to zero temperature $m=2m_n$ case. However, as observed in Fig.~\ref{Fig:EOS}, presence of temperature has a relative softening effect.
\begin{figure}[t]
\centering
\includegraphics[width=0.9\linewidth]{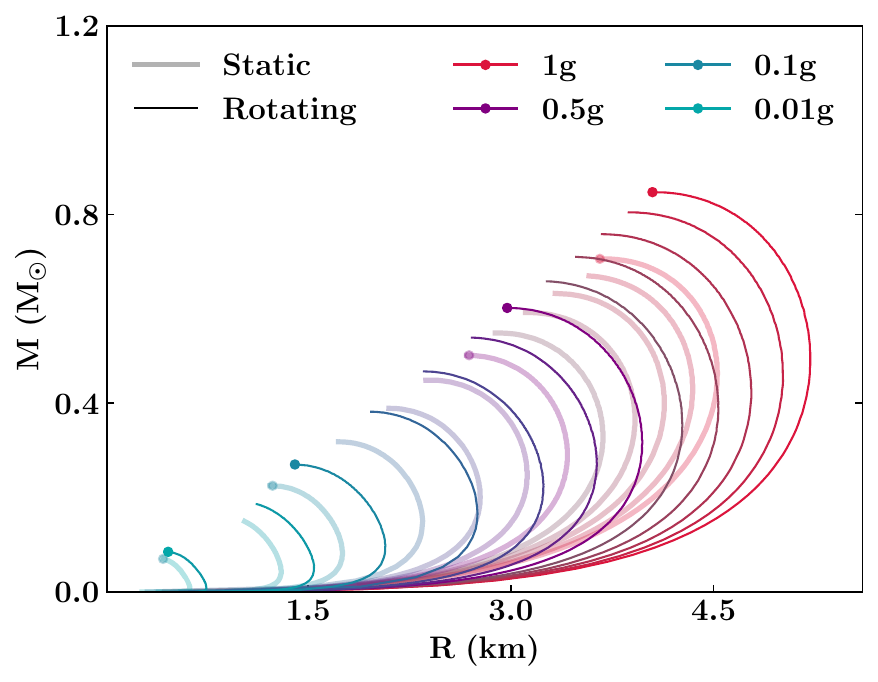}
\DeclareGraphicsExtensions{.pdf,.jpg}
\caption{Mass of the static and rotating BEC stars as a function of radius for $T=4\times 10^{11}$ K by varying the interaction strength $g$. }
\label{Fig:varied_g}
\end{figure}
\par
In Fig. \ref{Fig:varied_mn}, we present the mass-radius relation by varying the value of $m$, for $T=0$ and $T=4T_{11}$ cases.  Evidently the mass and radius of the star depend quite sensitively on the condensate mass $m$ considered. At zero temperature, on decreasing the mass of the condensate from $m = 1.5m_n$ to $m=1.25m_n$, the maximum mass and corresponding radius increased from $1.10 M_{\odot} $ to $1.44 M_{\odot}$ and $6.12$ km to $8.04$ km, for the static system
and $ 1.31 M_{\odot}$ to $1.72 M_{\odot} $ and $6.93$ km to $9.06$ km,
for the rotating case, respectively.
Also plotted for the reference are the previously considered $m=2m_n$ cases. Similar trend is observed for the finite temperature case. For $T = 4T_{11}$ case, the maximum mass and radius increased from $ 1.08 M_{\odot} $ to $ 1.42 M_{\odot} $ and $5.58$ km to $7.24$ km,
for the static case, and $ 1.29 M_{\odot} $ to $  1.70 M_{\odot}$ and $6.10$ km to $7.89$ km,
 for the rotating case.
The  central density at which the maximum mass achieved for $m = 1.5m_n$  is $\rho_c = 8.55 \times 10^{15}$ g/cm$^3$ and $\rho_c = 6.79 \times 10^{15}$ g/cm$^3$ for static and rotating system respectively at zero temperature. At $T=4T_{11},$ the  central density at which the maximum mass obtained is $\rho_c = 9.17 \times 10^{15}$ g/cm$^3$ and $\rho_c = 7.78 \times 10^{15}$ g/cm$^3$ for static and rotating system respectively.
The  central density at which the maximum mass achieved for $m = 1.25m_n$  is $\rho_c = 4.95 \times 10^{15}(5.43\times 10^{15})$ g/cm$^3$ and $\rho_c = 4  \times 10^{15} (4.60  \times 10^{15})$ g/cm$^3$ for the static and rotating system respectively at $T=0\, (4T_{11})$. The zero temperature results obtained above are in agreement with that of Ref.~\cite{Chavanis:2011cz}.
As observed earlier,
when the $m$ value decreases, EoS becomes more and more stiffer and stiffer EoS is known to exhibit higher mass and radius. When temperature is included a relative softening of EoS happens resulting in corresponding small decrease in the  maximum mass. We also observe that, as the value of $m$ gets decreased, although the difference between maximum masses obtained at each temperatures does not vary significantly for both the static and rotating cases, the difference between the corresponding radii keep on increasing.
\par
Finally, we proceed to study the effect of interaction between the bosons on the stellar profiles. We change the coupling constant of interaction $g\varpropto a/m$
in the EoS and solve the stellar structure equations for both static and rotating cases at finite temperature.
By varying the value of the interaction strength from $g=4.17 \times 10^{-43}$ g cm$^5$/s$^2$ to $0.5g$, $0.1g$ and $0.01g$, the maximum mass is observed to be decreasing from
$0.71 M_{\odot}$ to $0.50M_{\odot}$, 0.23$M_{\odot}$ and 0.07$M_{\odot}$ ($0.85 M_{\odot}$ to $0.60 M_{\odot}$, 0.27$M_{\odot}$ and 0.08$M_{\odot}$) respectively for the static (rotating) case at $T=4T_{11}$. This gradual decrease in maximum mass is depicted in Fig.~\ref{Fig:varied_g}.
We note that the mass-radius curves are highly sensitive to the interaction strength and reduction of the same results in smaller maximum masses.
As shown before, inclusion of rotation increases
the mass of the system relatively for any given value of $g$.
The self-interaction arising  due to the repulsive interaction between the individual particles within the condensate creates an effective outward pressure that counterbalances the gravitational collapse. If the self-interaction strength decreases, the effective pressure diminishes, and the force of gravity becomes dominant. As a result, the maximum mass that the BEC star can sustain decreases, as expected. This inference can be drawn from Eq.~(\ref{eos}), when self interaction is turned off the total outward pressure of the system against gravity vanishes, leading to an unstable configuration. This expected observation is in agreement with Ref.~\cite{Gruber:2014mna}.
Moreover, as we decrease the strength of self interaction, the difference between maximum masses attained for the static and rotating cases diminishes for any temperature.

\section{Summary and conclusions}
\label{summary}

We have investigated the global properties of static and slowly rotating Bose-Einstein condensate stars at finite temperature. To incorporate the effect of temperature in the analysis, we have used the recently developed finite temperature BEC EoS for the stellar matter, derived from the generalized Gross-Pitaevskii equation for the condensate and thermal fluctuations.
Gravity is treated in the framework of general relativity by using the Hartle-Thorne approximation for rotation.

Our numerical analysis shows that, although the impact of temperature considered is minimal on the EoS of BEC star, it is observed that
mass and radius of static and rotating BEC stellar configurations are highly sensitive to the temperature of the system.
The effect of increment of temperature is to reduce the mass and radius of the system;
however at lower central densities finite temperature supports higher values of mass.  Further, the introduction of rotation results in higher mass-radius stellar systems. Our analysis indicate that ignoring the general relativistic treatment of rotation results in erroneous estimation of global properties of BEC stars.
\par
Interestingly, we have found that presence of thermal fluctuations has negligible impact on maximum masses of the static as well as rotating BEC stars; although considerable change in maximum radii are seen.
We have also studied the effect of different rotational frequency on the system.
Further, together with the interplay of temperature and rotation, we have  analysed the effect of various EoS parameters, namely condensate mass and self-interaction strength, on the BEC star and quantified the results.
\par
The agreement between our results and that of non-rotating general relativistic BEC star at finite temperature considered in Ref. \cite{Latifah:2014ima} suggests that the maximum mass of the star remains unaffected by changes in temperature even in the rotating case. On the other hand, their utilization of a different EoS leads them to anticipate a distinct trend in the mass-radius behaviour.
Considering the difference in their theoretical approach, leading to stiffer EoS with the increasing temperatures, a close comparison with our present work is irrelevant. Furthermore, inclusion of the effect of magnetic field in the static finite temperature BEC star also reports minimal effect on the maximum mass \cite{Angulo:2022gpj}.
In future, we would like to extend our studies by incorporating magnetic field into the finite temperature BEC star system. Also, the possibility of rotating neutron stars with different exotic matter with such BEC at its core will be of interest.

\section*{Acknowledgements}
Authors would like to acknowledge discussions with Axel Pelster, Jitesh R. Bhatt and Sandeep Gautam.
PSA and PSK would like to thank the warm hospitality of IIT Ropar, where part of this work was done.
LJN  acknowledges the Department of Science and Technology, Government of India for the
INSPIRE Fellowship.

\bibliography{references.bib}{}

\end{document}